\documentclass[aps,showkeys,showpacs,notitlepage,superscriptaddress]{revtex4-1}
\usepackage{amsmath}
\usepackage{amssymb}
\usepackage{graphicx}
\usepackage{hyperref}
\usepackage{color}
\usepackage[utf8]{inputenc}
\hypersetup{pdftitle=Forman curvature for complex networks}

\begin{document}
\title{Forman curvature for complex networks}
\author{R.P. Sreejith}
\affiliation{The Institute of Mathematical Sciences, Chennai, India}
\author{Karthikeyan Mohanraj}
\affiliation{The Institute of Mathematical Sciences, Chennai, India}
\author{J\"urgen Jost}
\email{jost@mis.mpg.de}
\affiliation{Max Planck Institute for Mathematics in the Sciences, Leipzig, Germany}
\affiliation{The Santa Fe Institute, Santa Fe, New Mexico, USA}
\author{Emil Saucan}
\email{emil.saucan@mis.mpg.de}
\affiliation{Max Planck Institute for Mathematics in the Sciences, Leipzig, Germany}
\affiliation{Department of Electrical Engineering, Technion, Israel Institute of Technology, Haifa, Israel}
\author{Areejit Samal}
\email{asamal@imsc.res.in}
\affiliation{The Institute of Mathematical Sciences, Chennai, India}
\begin{abstract}
We adapt Forman's discretization of Ricci curvature to the case of undirected networks, both weighted and unweighted, and investigate the measure in a variety of model and real-world networks. We find that most nodes and edges in model and real networks have a negative curvature. Furthermore, the distribution of Forman curvature of nodes and edges is narrow in random and small-world networks, while the distribution is broad in scale-free and real-world networks. In most networks, Forman curvature is found to display significant negative correlation with degree and centrality measures. However, Forman curvature is uncorrelated with clustering coefficient in most networks. Importantly, we find that both model and real networks are vulnerable to targeted deletion of nodes with highly negative Forman curvature. Our results suggest that Forman curvature can be employed to gain novel insights on the organization of complex networks.
\end{abstract}
\maketitle

\section{Introduction}
\label{introduction}

Complex networks \cite{Watts1998,Barabasi1999,Albert2002,Newman2010,Fortunato2010,Dorogovtsev2013} pervade our everyday life. These range from biological networks \cite{Barabasi2004} such as metabolic network \cite{Jeong2000} representing interactions among metabolites and enzymes responsible for growth and maintenance of a cell, to transportation networks \cite{Subelj2011} such as the road, rail and airline network linking cities across the world, to social networks such as Facebook \cite{Ellison2007} linking individuals based on their friendships. A key goal of network theory is to characterize the structure of networks and investigate structure-function relationships \cite{Watts1998,Barabasi1999,Albert2002,Newman2010,Fortunato2010,Dorogovtsev2013}.

In this direction, a focus has been the geometrical characterization of model and real-world networks \cite{Eckmann2002,Ollivier2009,Lin2010,Lin2011,Bauer2012,Jost2014,Wu2015,Ni2015,Sandhu2015a}.
In geometry, curvature plays a central role, since it represents a measure to quantify the deviation of a geometrical object from being flat. Given the essential role of curvature in differential geometry wherefrom this notion originated, the search for its analogue in various generalized settings, and in particular, in the discrete case is only natural. However, only with the advent of computer science and modern graph theory did this drive attend maturity (see below).

Sectional curvature is the most expressive classical curvature, that has proved to be extremely fruitful leading to significant generalizations in geometry (see, for example \cite{Burago2001,Plaut2001} and references therein). However, the sectional curvature with the notable exception of combinatorial curvature (or angular defect), has turned out to be less fruitful in the context of graph theory. Combinatorial curvature has led to several applications in graph theory, computational geometry, graphics and imaging. In particular, the role of combinatorial curvature in studying the properties of various types of networks has been realized and exploited, even if, in many cases, only in the simplest form of clustering coefficient \cite{Eckmann2002}. Shortly, a number of other more elaborate and \textit{geometric} types of sectional curvature have also been considered \cite{Shavitt2004,Saucan2005,Narayan2011,Wu2015}.

In contrast to sectional curvature, Ricci curvature is a more abstract and less intuitive concept. The work of Grigori Perelman on the geometrization conjecture \cite{Perelman2002,Perelman2003} has led to renewed interest in Ricci curvature and its discretizations. Ricci curvature has proved to be more versatile and flexible compared to sectional curvature, rendering many discretizations for graphs and alike structures  \cite{Stone1976,Bakry1985,Chow2003,Morgan2005,Jin2007,Bonciocat2009,Lott2009,Alsing2011,Gu2013}.

In particular, Ollivier's discretization of the Ricci curvature \cite{Ollivier2009,Ollivier2010,Ollivier2013} has proven to be successful for both conceptual advances \cite{Lin2010,Lin2011,Bauer2012,Jost2014,Loisel2014} and practical applications \cite{Ni2015,Sandhu2015a,Sandhu2015b} in the domain of networks. However, Ollivier-Ricci curvature has some limitations. Ollivier-Ricci curvature necessitates the calculation of the earth mover's distance or Wasserstein 1-metric, and this in practice requires solving a linear programming problem. Thus, the calculation of Ollivier-Ricci curvature is computationally intensive and may be forbidden for very large networks. Ollivier-Ricci curvature has formal basis in optimal transport theory, and thus, Ollivier's curvature is well suited to investigate information transfer in communication networks. However, Ollivier-Ricci curvature may be less suited to investigate interaction networks such as protein-protein interactions.

In this paper, we introduce another discretization of the classical Ricci curvature proposed by R. Forman \cite{Forman2003} to the domain of complex networks. Remarkably, Forman's curvature for complex networks is enticingly simple to compute in large-scale networks. Forman's curvature is general and flexible which renders it suitable for most situations. Notably, unlike clustering coefficient, Forman's curvature is defined for any edge in the network, and does not presume the existence of triads or triangles in the graph. Moreover, the definition of Forman's curvature elegantly incorporates the weights of edges and nodes in the network. Hence, Forman's curvature is suitable to study both unweighted and weighted networks.

The remainder of this paper is structured as follows. In section \ref{FormanCurvature}, we introduce the Forman curvature for networks. In section \ref{dataset}, we describe the model and real networks used to investigate the Forman curvature for networks, while section \ref{results} contains our main results. We conclude, in section \ref{conclusion}, with suggestions for future extensions and possible applications.


\section{Definition of Forman curvature for networks}
\label{FormanCurvature}

\begin{figure}
\includegraphics[width=.43\columnwidth]{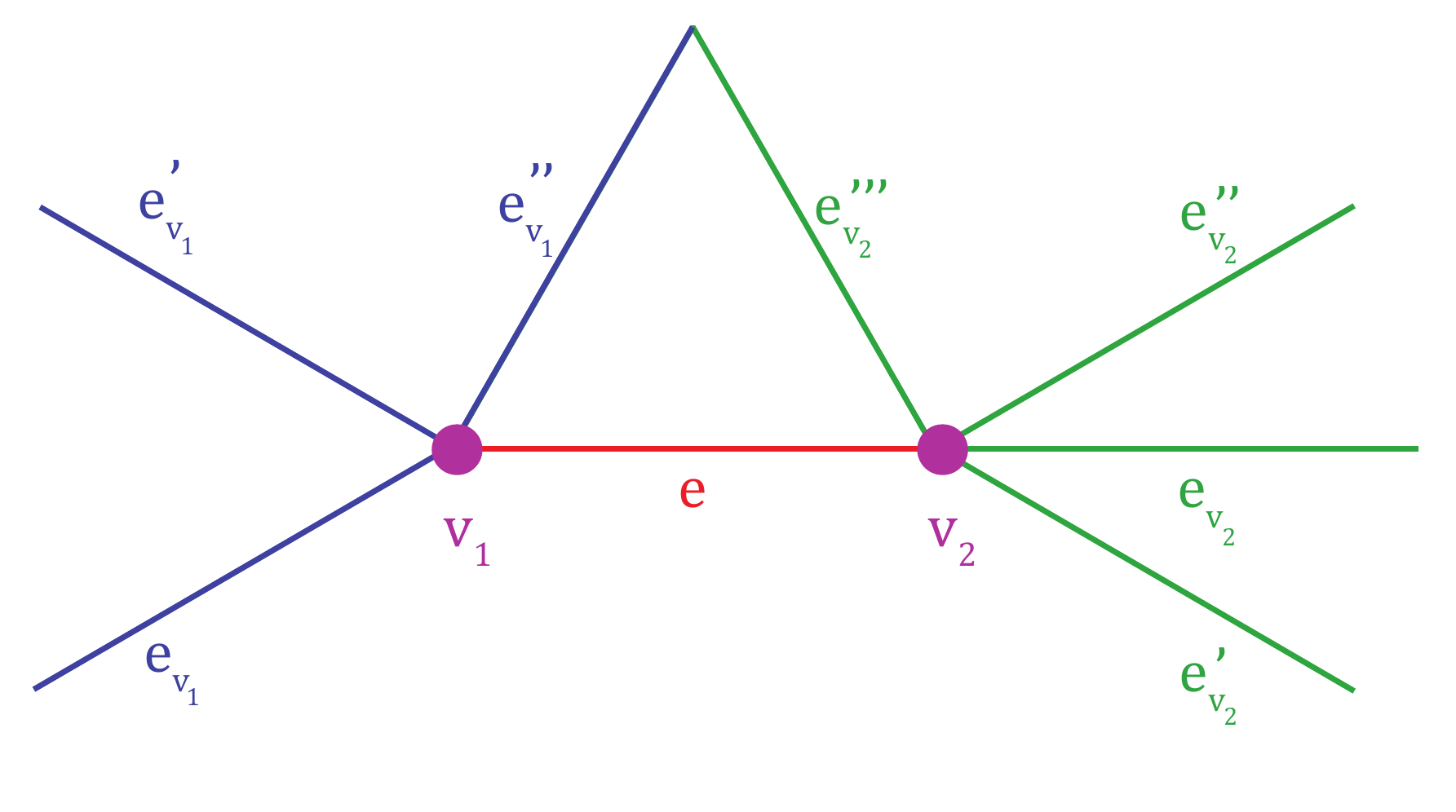}
\caption{Forman curvature of edge $e$ connecting the nodes $v_1$ and $v_2$ and contribution from edges parallel to the edge $e$ under consideration. An edge is considered to be \textit{parallel} to a given edge $e$, if it has in common with $e$ either a \textit{child} (i.e., a lower dimensional face), or a \textit{parent} (i.e., a higher dimensional face), but not both simultaneously. The edges $e_{v_1}$, $e_{v_1}^{\prime}$, $e_{v_1}^{\prime\prime}$ are parallel to $e$ because they share the node $v_1$. Analogously, the edges $e_{v_2}$, $e_{v_2}^{\prime}$, $e_{v_2}^{\prime\prime}$, $e_{v_2}^{\prime\prime\prime}$ are parallel to $e$ because they share the node $v_2$. See Appendix for additional discussion.}
\label{edge_parallel}
\end{figure}

In this section, we introduce Forman curvature to the realm of complex networks, and present the mathematical formula to compute the Forman curvature in undirected networks.

Forman's \cite{Forman2003} discretization of the classical Ricci curvature holds for a broad class of geometric objects, the so called (\textit{regular}) $CW$ \textit{complexes} \cite{Whitehead1949a,Whitehead1949b}. A full definition of this notion is out of the scope of this paper, and the interested reader is directed to Forman's article \cite{Forman2003} for details. Here, we only wish to emphasize that the broad class of geometric objects for which Forman's discretization of the classical Ricci curvature holds include polyhedra, triangular meshes, and most important for this work, (locally finite) graphs.

It is worthwhile to mention that Forman's definition of the classical Ricci curvature for the case of cell complexes is based on a proper interpretation of the so called \textit{Bochner-Weitzenb\"{o}ck formula}. Again the mathematical expression of the Forman-Ricci curvature is extremely technical and less relevant in the present context of networks, and the interested reader may refer for example \cite{Jost2011} for details. We remark that, in the classical context, the Forman Ricci curvature gives the relationship between the Riemannian \textit{good} Laplacian and the so-called \textit{rough} Laplacian, in terms of a complicated expression containing the curvatures of the manifold. Simply stated, it corrects the standard Laplacian which is also known as the Laplace-Beltrami operator, in terms of the curvatures of the underlying manifold. Since Laplacian plays a key role in heat equation (see, e.g. \cite{Jost2011}), one can gain some physical intuition by realizing that the heat evolution on a curved metal plate differs from that on a planar plate in a manner that is evidently dependent on the shape (i.e., curvature) of the plate. We also suggest possible applications of the Forman's version of the Laplacian in our concluding remarks.

Due to the technical nature of the Forman-Ricci curvature, we choose not to present here the expression for Forman's curvature in its general form, but would rather restrict ourselves to highlighting the following important point. Ricci curvature operates directionally along vectors. A direct consequence of this fact is that, in our discrete setting, Forman's curvature is associated with the discrete analogue of a vector, namely, edges.

Forman curvature for an edge (which is in dimension 1) is given by the following formula:
\begin{widetext}
\begin{equation}
\label{FormanRicciEdge}
\mathbf{F}(e) = w_e \left( \frac{w_{v_1}}{w_e} +  \frac{w_{v_2}}{w_e}  - \sum_{e_{v_1}\ \sim\ e,\ e_{v_2}\ \sim\ e} \left[\frac{w_{v_1}}{\sqrt{w_e w_{e_{v_1} }}} + \frac{w_{v_2}}{\sqrt{w_e w_{e_{v_2} }}} \right] \right)\,
\end{equation}
\end{widetext}
where
\begin{itemize}
\item $e$ denotes the edge under consideration between two nodes $v_1$ and $v_2$.
\item $w_e$ denotes the weight of the edge $e$ under consideration.
\item $w_{v_1}$ and $w_{v_2}$ denote the weights associated with the nodes $v_1$ and $v_2$, respectively.
\item $e_{v_1} \sim e$ and $e_{v_2} \sim e$ denote the set of edges incident on nodes $v_1$ and $v_2$, respectively, after \textit{excluding} the edge $e$ under consideration which connects the two nodes  $v_1$ and $v_2$ (see Figure \ref{edge_parallel}).
\end{itemize}
Note that \textit{by definition} this notion of curvature is intrinsically associated with edges, and thus, Forman curvature is inherently suited to networks. Also, this notion of curvature does not necessitate any artifice such as extending a measure for the curvature of nodes to the edges.

On the other hand, while Forman curvature is essentially defined on edges (as it represents a discretization of Ricci curvature), it can be elegantly extended to nodes as follows:
\begin{equation}
\label{FormanRicciNode}
\mathbf{F}(v) = \frac{1}{\text{deg}(v)}\sum_{e_v\ \sim\ v} \mathbf{F}(e_v) \,
\end{equation}
where $e_v$ denotes the set of edges incident on the node $v$ and deg($v$) denotes the degree of node $v$. This definition of the Forman curvature of nodes in network is handy because several standard measures (such as degree and clustering coefficient) to characterize the structure of networks are defined for the nodes. Thus, the definition of the Forman curvature of nodes enables a comparative analysis with other node based network measures.

A full explanation of the formula above, in particular of the role of parallel edges, is beyond the scope of the present paper, and we refer the interested reader to Forman's original article \citep{Forman2003}. However, we provide an intuitive interpretation of Forman curvature in the Appendix.


\section{Dataset of model and real networks}
\label{dataset}

We have analyzed the Forman curvature in diverse networks which include both model and real-world networks. In this study, we have considered the following generative models for networks:
\begin{itemize}
\item \textbf{Erd\"{o}s-R\'{e}nyi (ER) model} \cite{Erdos1961} is widely used to generate random graphs. ER model gives an ensemble $G(n,p)$ of graphs where $n$ is the number of nodes and $p$ is the probability that each possible edge exists between any pair of nodes in the network.
\item \textbf{Watts-Strogatz (WS) model} \cite{Watts1998} generates graphs with small-world properties of high clustering coefficient and small average path length. This model initially starts with a regular ring lattice, a graph with $n$ nodes where each node is connected to its $k$ nearest neighbors. Next an endpoint of each edge in the regular ring lattice is rewired with probability $\beta$ to a new node (which is selected from all possible nodes in the network with a uniform probability).
\item \textbf{Barab\'{a}si-Albert (BA) model} \cite{Barabasi1999} generates scale-free networks with power-law degree distribution. This growing network model initially starts with a graph of $m_0$ nodes. Next new nodes are added to the initial graph one at a time, and each new node is connected to $m$ $\le$ $m_0$ existing nodes with a probability that is proportional to the degree of existing nodes. Formally, in this preferential attachment model, the probability $p_i$ that the new node is connected to existing node $i$ is given by $$p_i = \frac{k_i}{\sum_j k_j},$$ where $k_i$ is the degree of existing node $i$ and the sum in the denominator is taken over the degree of all existing nodes $j$. Thus, high-degree nodes acquire more edges over time compared to low-degree nodes in this model.
\item \textbf{Power-law cluster (PLC) model} \cite{Holme2002} is an extension of the BA model that generates scale-free networks with power-law degree distribution and approximate average clustering. Compared to the BA model, the PLC model contains an extra step where the addition of a random edge between a new node and an exiting node $i$ is followed by the possibility of adding an edge between the new node and a neighbor of the existing node $i$. This extra step in the PLC model renders the formation of triads more likely, and thus, can generate networks with high average clustering coefficient.
\end{itemize}

In addition to the model networks, we have analyzed the following undirected real-world networks:
\begin{itemize}
\item \textbf{US Power Grid network} \cite{Leskovec2007} captures the high-voltage power grid in the western states of USA. In this network of 4941 nodes and 6594 edges, the nodes are generators or transformers or substations, and the edges are power supply lines.
\item \textbf{Euro road network} \cite{Subelj2011} represents the international E-road network which is a road network located mostly in Europe. In this network of 1174 nodes and 1417 edges, the nodes are cities, and the edges are roads linking them.
\item \textbf{PGP network} \cite{Boguna2004} represents the interaction network of users of the Pretty Good Privacy (PGP) algorithm. In this network of 10680 nodes and 24316 edges, the nodes are users of Pretty Good Privacy (PGP) algorithm, and edges are interactions between the users.
\item \textbf{Email communication network} \cite{Guimera2003} represents the Email communication at the University Rovira i Virgili in Tarragona in the south of Catalonia in Spain. In this network of 1133 nodes and 5451 edges, the nodes are users, and the edges represent direct communication between users.
\item \textbf{Yeast protein interactions} \cite{Jeong2001} represents the interactions between proteins of yeast \textit{Saccharomyces cerevisiae}. In this network of 1870 nodes and 2277 edges, the nodes are proteins, and the edges are interactions between proteins.
\item \textbf{PDZ domain interactions} \cite{Beuming2005} represents the network of protein–protein interactions obtained from PDZbase. In this network of 212 nodes and 244 edges, the nodes are proteins, and the edges are PDZ-domain mediated interactions between proteins.
\item \textbf{Adjective-Noun adjacency network} \cite{Newman2006} gives the common noun and adjective adjacencies for the novel David Copperfield written by Charles Dickens. In this network of 112 nodes and 425 edges, the nodes are nouns or adjectives, and the edges represent their presence in adjacent positions in the novel.
\item \textbf{Jazz musicians network} \cite{Gleiser2003} represents the collaboration network between Jazz musicians. In this network of 198 nodes and 2742 edges, the nodes are Jazz musicians, and the edges represent collaboration between musicians.
\item \textbf{Facebook network} \cite{Mcauley2012} represents the user–user friendships in the online social network Facebook. In this network of 2888 nodes and 2981 edges, the nodes are users, and the edges represent friendship between users of Facebook.
\item \textbf{Hamsterster Friendship} represents the network of friendships and family links between users of the website hamsterster.com. In this network of 2426 nodes and 16631 edges, the nodes are users of hamsterster.com, and the edges represent friendship or family links between users.
\item \textbf{Gnutella} \cite{Leskovec2007} captures the Gnutella peer-to-peer file sharing network from August 2002. In this network of 6301 nodes and 20777 edges, the nodes are hosts in Gnutella network, and the edges are connections between hosts.
\item \textbf{Human protein interactions} \cite{Rual2005} represents the initial version of a proteome-scale map of human binary protein–protein interactions. In this network of 3133 nodes and 6726 edges, the nodes are proteins, and the edges are interactions between human proteins.
\item \textbf{Noun occurrences in Bible} \cite{Basu2016} represents the network of nouns (i.e., places and names) in the King James Bible and information about their occurrences. In this network of 1773 nodes and 9131 edges, the nodes are nouns, and the edges represent occurrences of nouns in a single verse in Bible. Note unlike other undirected networks considered here, this undirected network has positive edge weights.
\end{itemize}
Majority of the above-mentioned undirected real networks were downloaded from the KONECT \cite{Kunegis2013} database (cf. Supplementary Table S1).

We have analyzed the structure of the considered model and real networks based on several network measures such as maximum degree, minimum degree, average degree, number of connected components, size of the largest connected component, mean shortest path length, average clustering coefficient and degree assortativity, and these results are contained in Supplementary Tables S2-S4. Note that we take the weight of nodes and edges to be 1 while computing the Forman curvature in unweighted networks.

\begin{figure}[!]
\includegraphics[width=.34\columnwidth]{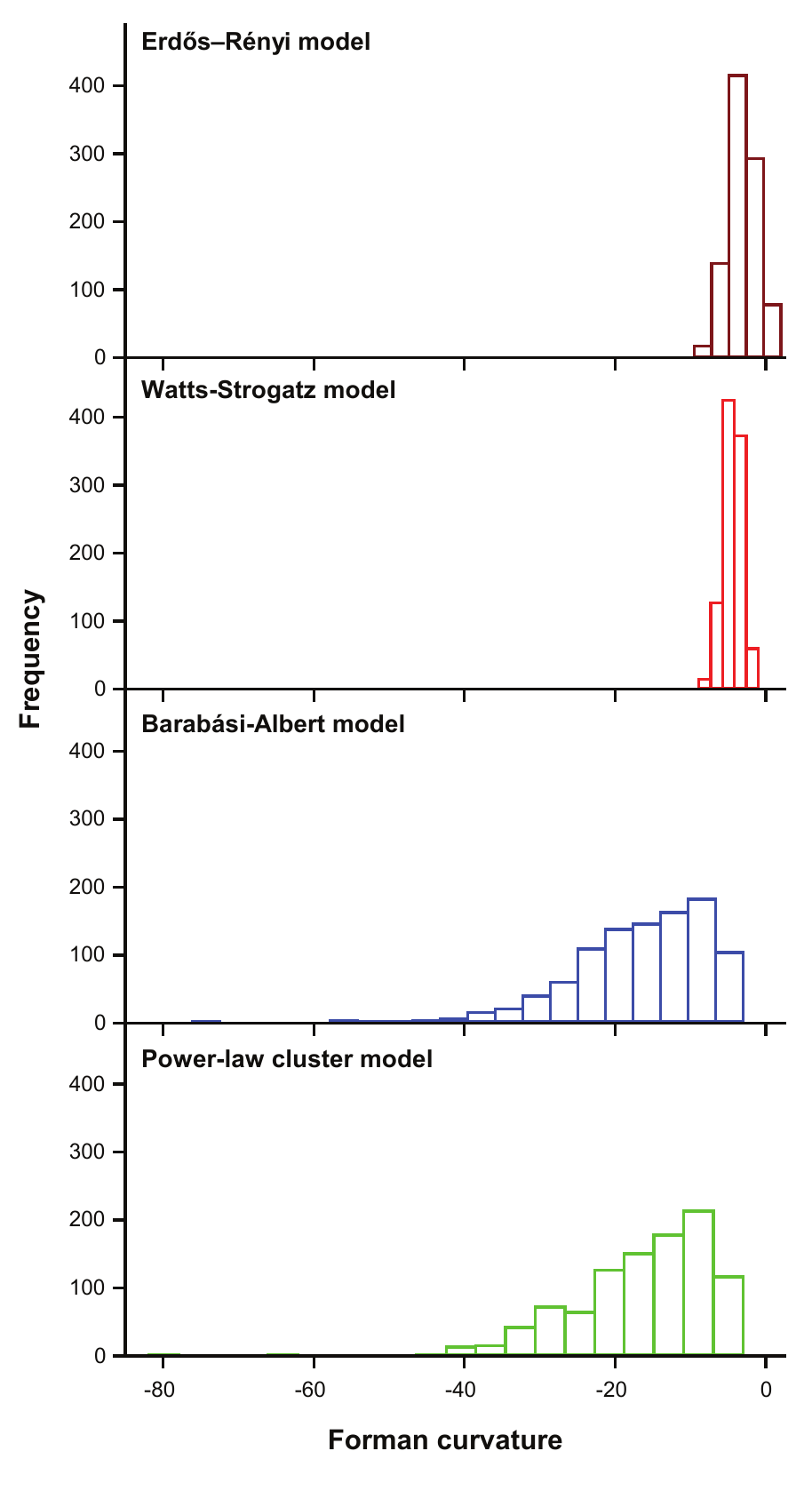}
\caption{Distribution of Forman curvature of nodes in model networks. Erd\"{o}s-R\'{e}nyi (ER) model with parameters: number of nodes $n=1000$ and probability $p$ that two nodes in the graph are directly connected $=0.003$. Watts-Strogratz (WS) model with parameters: number of nodes $n=1000$, each node in the initial ring topology is connected to its $k=5$ nearest neighbors, and probability $p$ of rewiring each edge $=0.5$. Barab\'{a}si-Albert (BA) model with parameters: number of nodes $n=1000$, and the number of random edges to be added to each new node $m=3$. Power-law cluster (PLC) model with parameters: number of nodes $n=1000$, the number of random edges to be added to each new node $m=3$, and probability $p$ of adding a triangle after adding a random edge $=0.05$.}
\label{dist_model}
\end{figure}

\begin{figure}[!]
\includegraphics[width=.34\columnwidth]{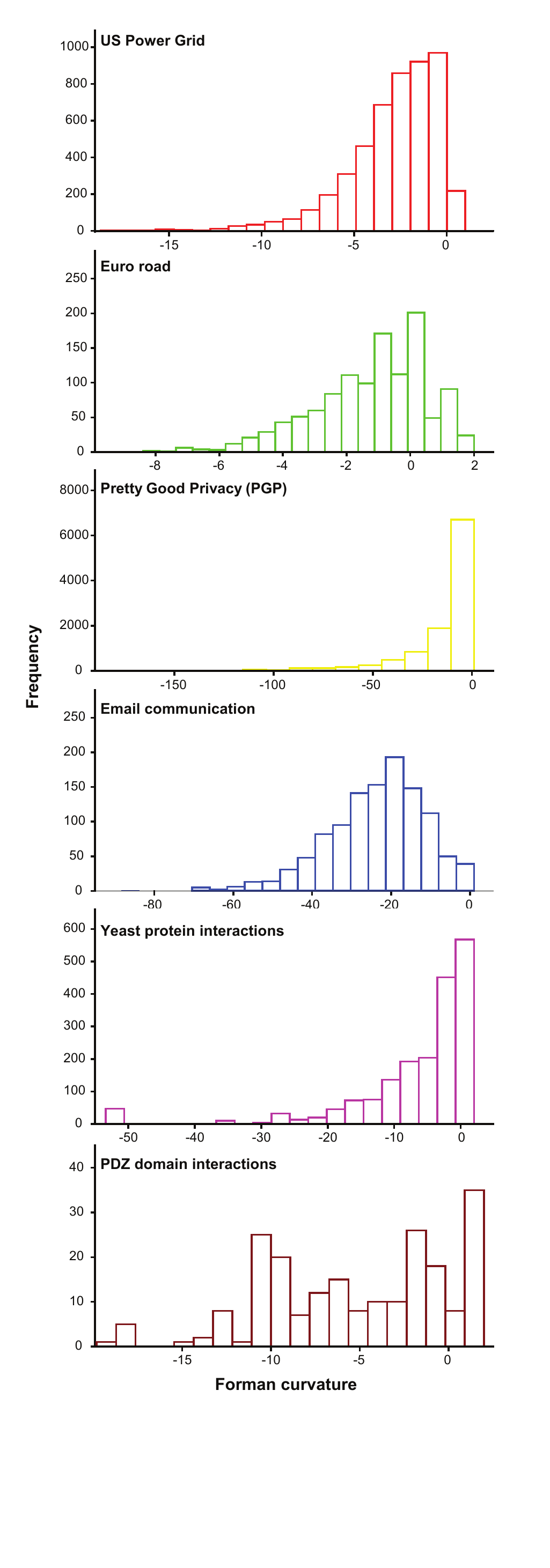}
\caption{Distribution of Forman curvature of nodes in real networks: US Power Grid, Euro road, Pretty Good Privacy (PGP), Email communication, Yeast protein interactions, and PDZ domain interactions.}
\label{dist_real}
\end{figure}

\begin{figure*}[!]
\includegraphics[width=12cm]{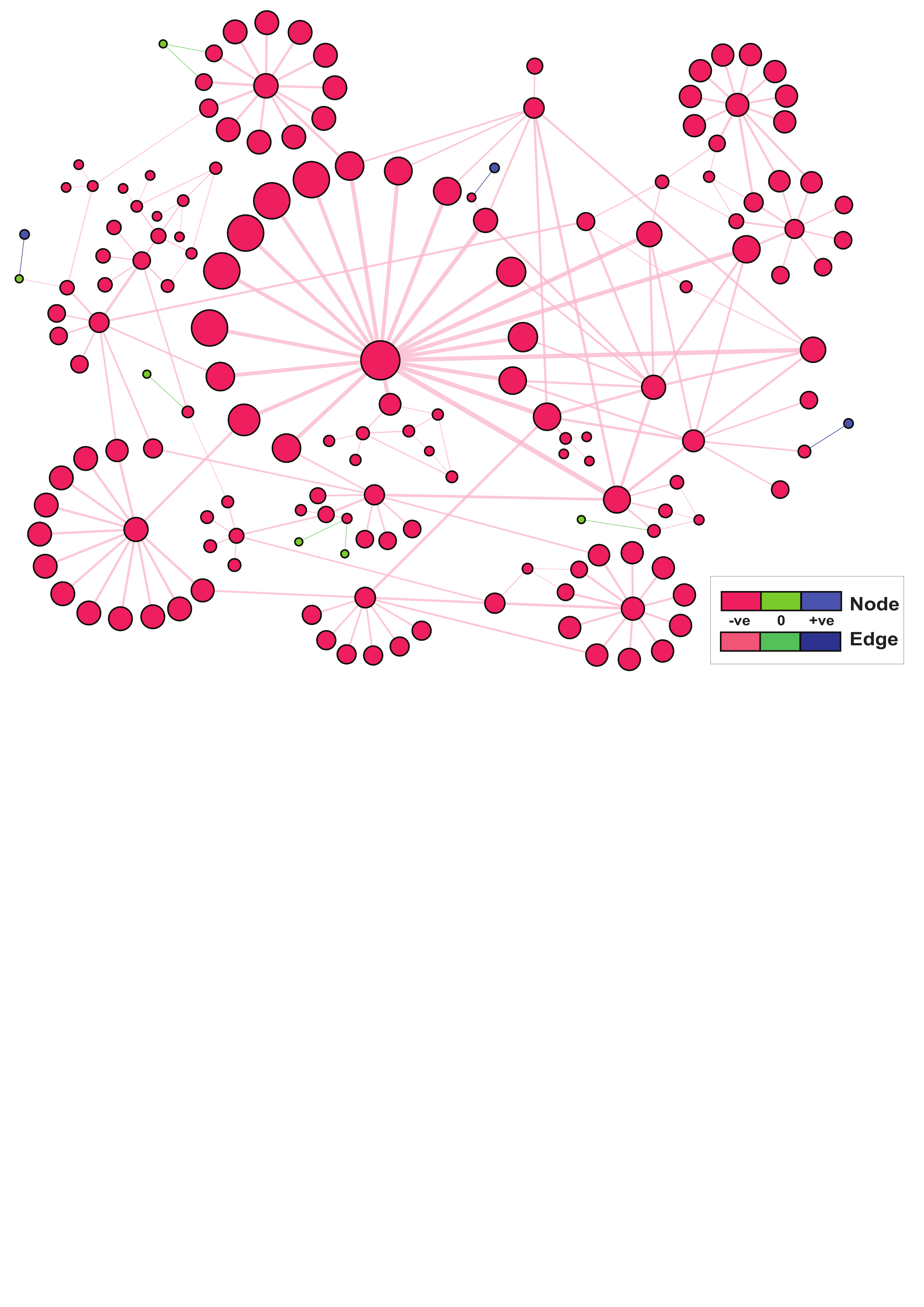}
\caption{Largest connected component of the PDZ-domain mediated protein-protein interaction network. Here, nodes are proteins and edges are interactions between them. Nodes and edges displayed in red, blue and green have negative curvature, zero curvature and positive curvature, respectively. Size of nodes and width of edges are proportional to the absolute value of the corresponding Forman curvature.}
\label{PDZ_network}
\end{figure*}

\begin{figure*}[!]
\includegraphics[width=12cm]{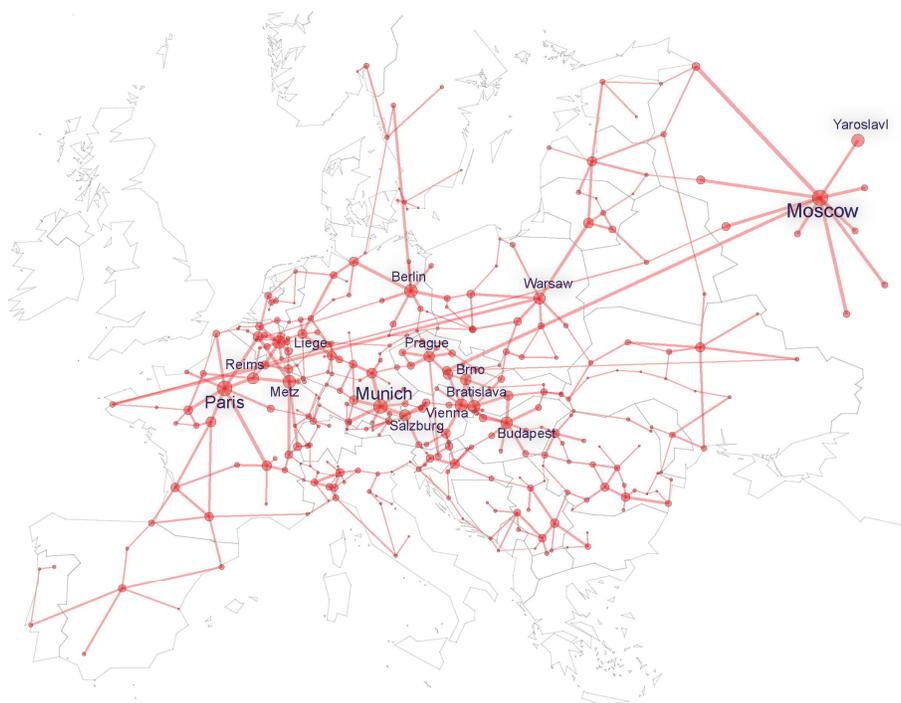}
\caption{Euro road network based on the international E-road network connecting cities located mostly in Europe. Here, nodes are cities and edges are roads links between the cities. In this figure, cities have been placed based on their geographical location (i.e., latitude and longitude) on the map of the world. Here, we show a section of the Euro road network limited to the cities with Forman curvature of nodes $\le -2$ and edges between them. Size of nodes and width of edges are proportional to the absolute value of the corresponding Forman curvature. Cities with Forman curvature $<-7$ are labeled here.}
\label{Euroroad_network}
\end{figure*}

\begin{figure}[!]
\includegraphics[width=.5\columnwidth]{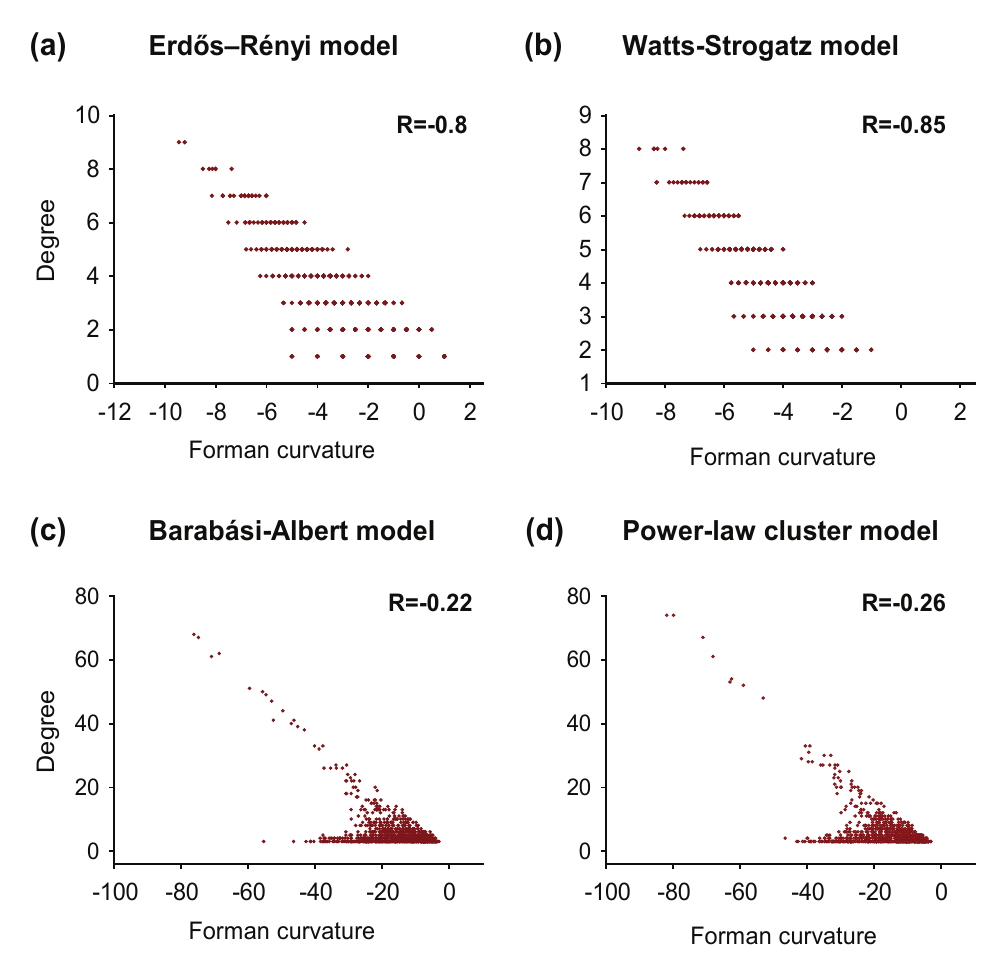}
\caption{Correlation between Forman curvature and degree of nodes in model networks. (a) Erd\"{o}s-R\'{e}nyi (ER) model. (b) Watts-Strogratz (WS) model. (c) Barab\'{a}si-Albert (BA) model. (d) Power-law cluster (PLC) model. The parameters used to construct graphs from the generative models are same as those mentioned in Figure \ref{dist_model}. In Figures \ref{cor_degree_model}-\ref{cor_bc_real}, we also indicate the Pearson correlation coefficient $R$ for each case.}
\label{cor_degree_model}
\end{figure}

\begin{figure}[!]
\includegraphics[width=.5\columnwidth]{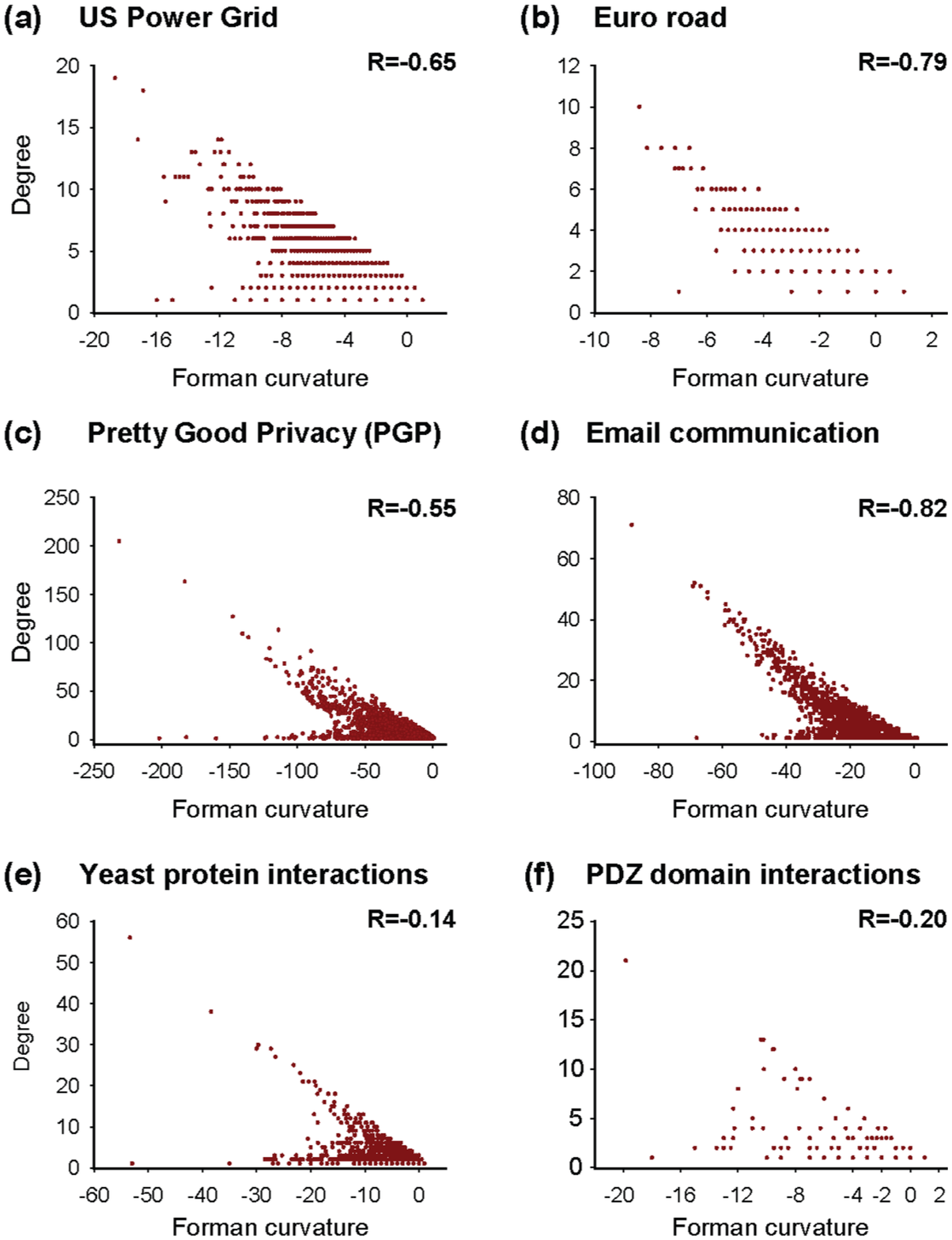}
\caption{Correlation between Forman curvature and degree of nodes in real networks. (a) US Power Grid. (b) Euro road. (c) Pretty Good Privacy (PGP). (d) Email communication. (e) Yeast protein interactions. (f) PDZ domain interactions.}
\label{cor_degree_real}
\end{figure}

\begin{figure}[!]
\includegraphics[width=.5\columnwidth]{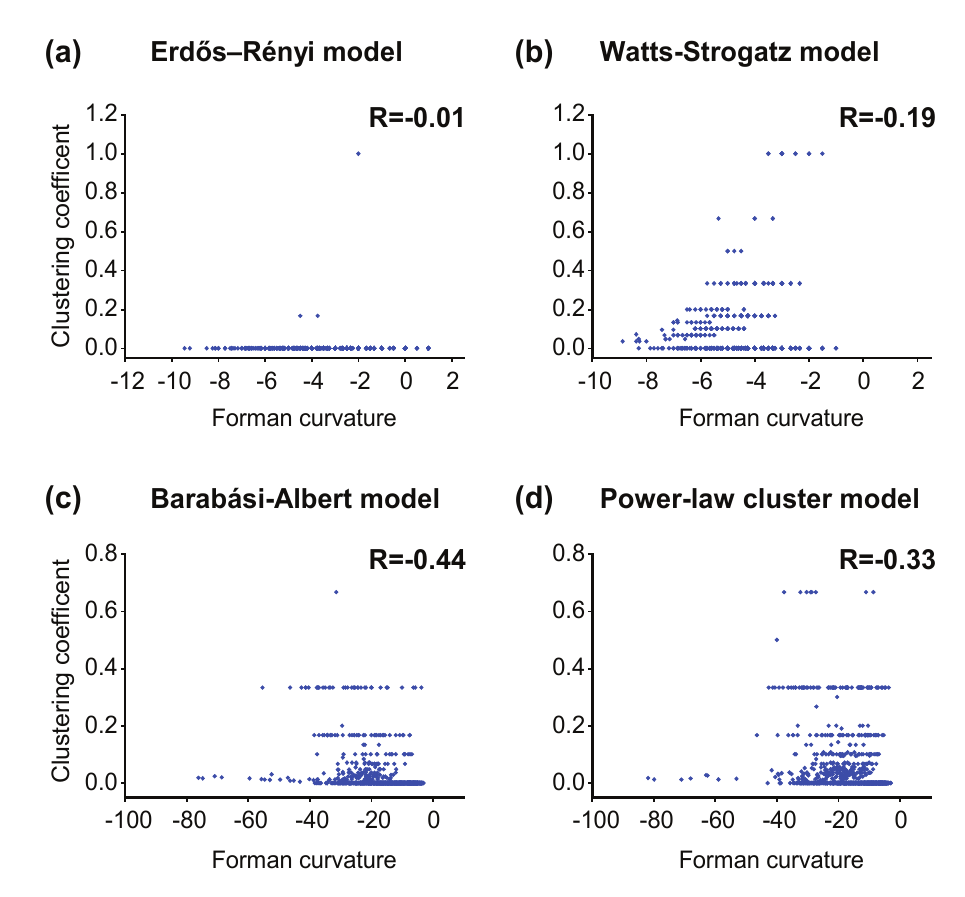}
\caption{Correlation between Forman curvature and clustering coefficient of nodes in model networks. (a) Erd\"{o}s-R\'{e}nyi (ER) model. (b) Watts-Strogratz (WS) model. (c) Barab\'{a}si-Albert (BA) model. (d) Power-law cluster (PLC) model. The parameters used to construct graphs from these generative models are same as those mentioned in Figure \ref{dist_model}.}
\label{cor_cc_model}
\end{figure}

\begin{figure}[!]
\includegraphics[width=.5\columnwidth]{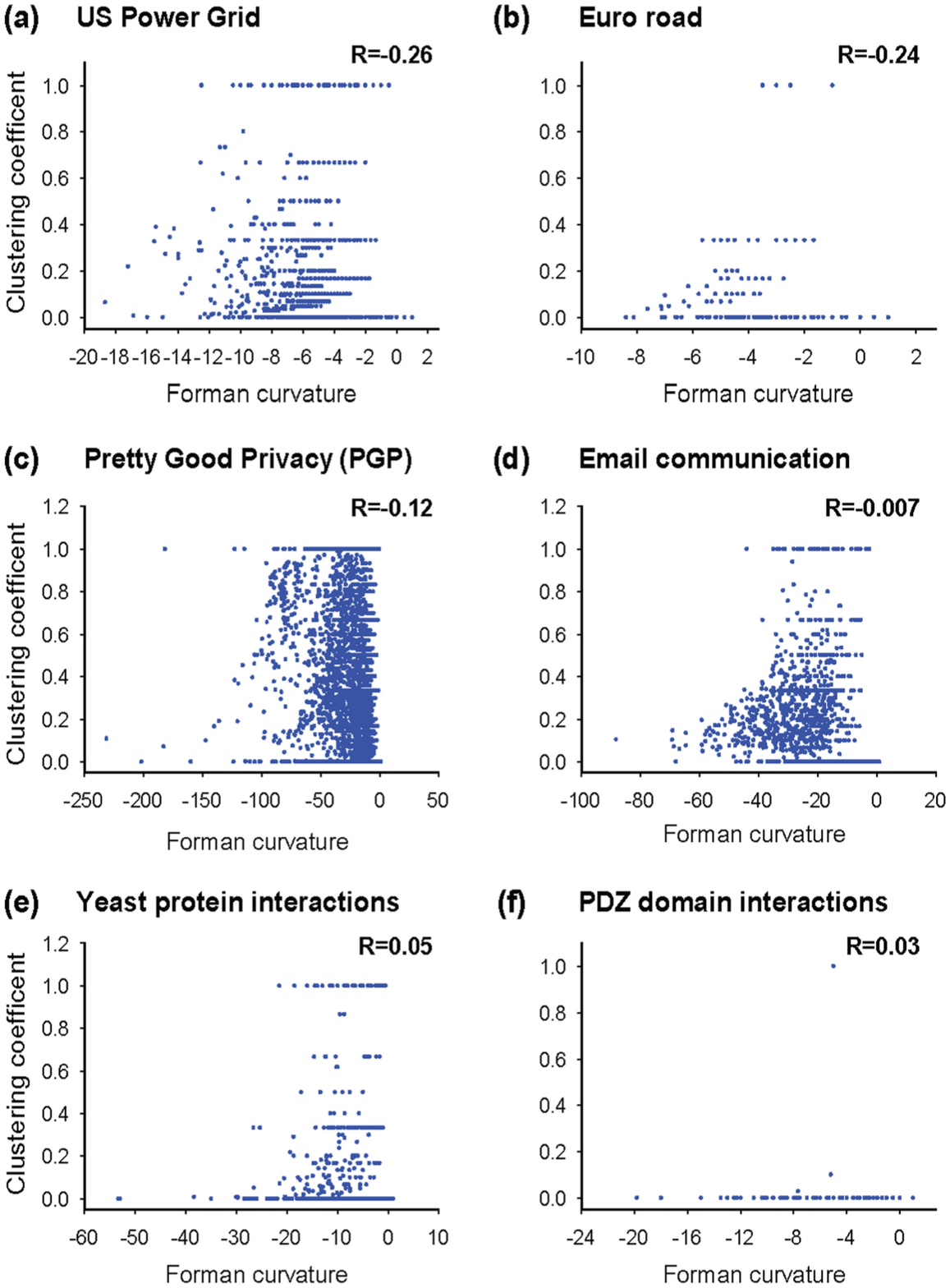}
\caption{Correlation between Forman curvature and clustering coefficient of nodes in real networks. (a) US Power Grid. (b) Euro road. (c) Pretty Good Privacy (PGP). (d) Email communication. (e) Yeast protein interactions. (f) PDZ domain interactions.}
\label{cor_cc_real}
\end{figure}

\begin{figure}[!]
\includegraphics[width=.5\columnwidth]{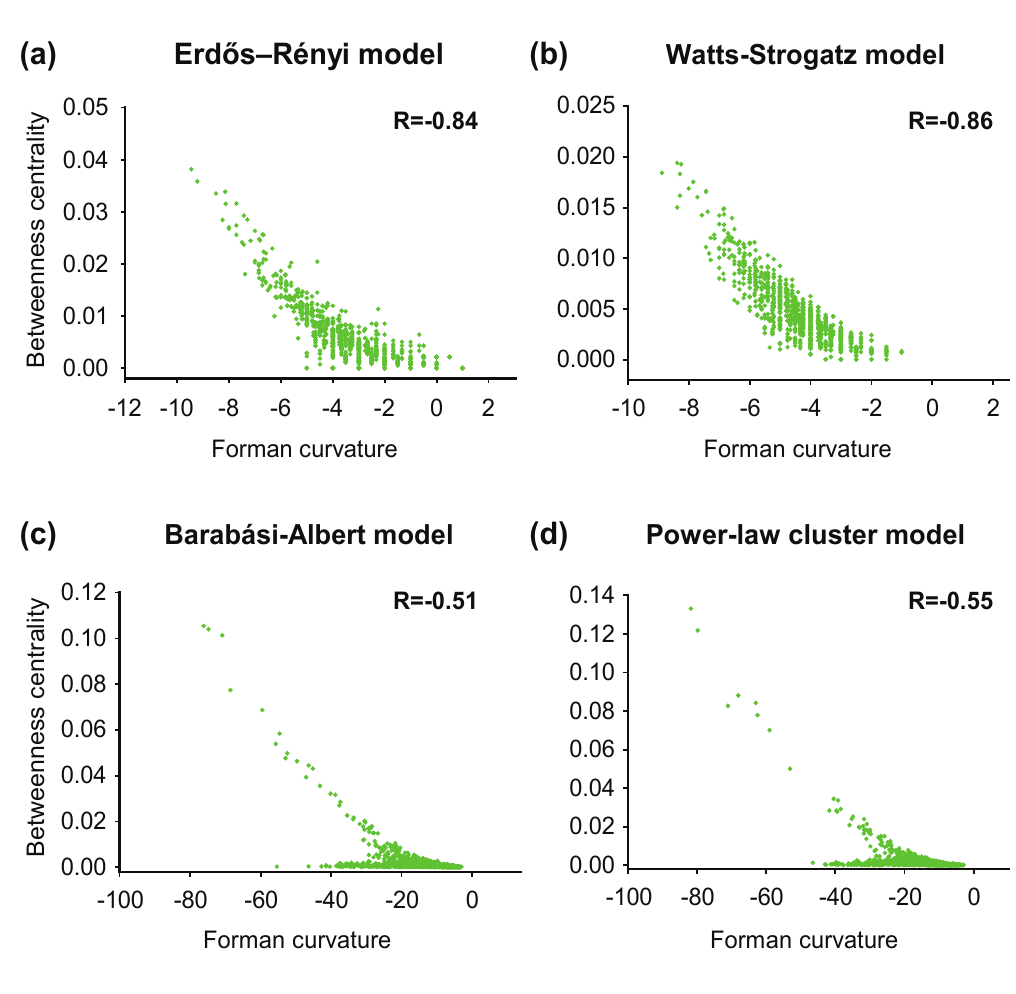}
\caption{Correlation between Forman curvature and betweenness centrality of nodes in model networks. (a) Erd\"{o}s-R\'{e}nyi (ER) model. (b) Watts-Strogratz (WS) model. (c) Barab\'{a}si-Albert (BA) model. (d) Power-law cluster (PLC) model. The parameters used to construct graphs from these generative models are same as those mentioned in Figure \ref{dist_model}.}
\label{cor_bc_model}
\end{figure}

\begin{figure}[!]
\includegraphics[width=.5\columnwidth]{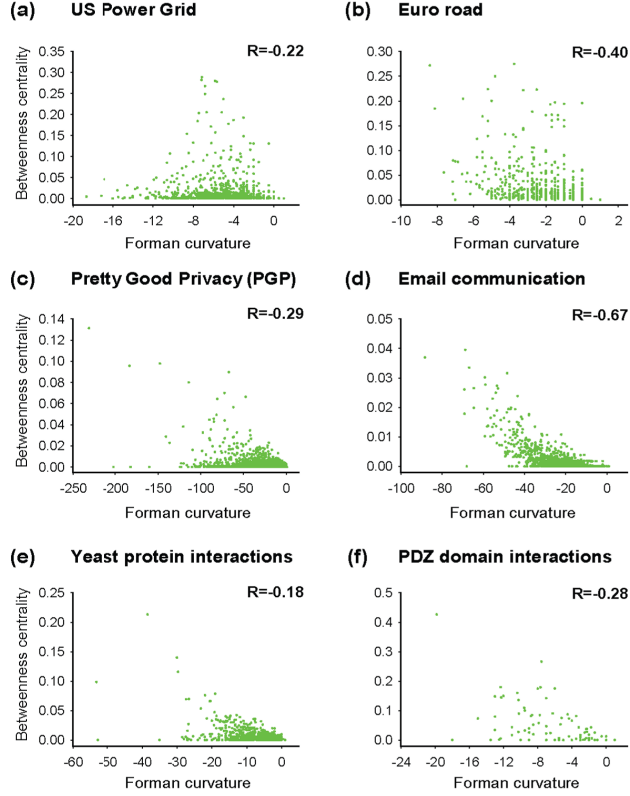}
\caption{Correlation between Forman curvature and betweenness centrality of nodes in real networks. (a) US Power Grid. (b) Euro road. (c) Pretty Good Privacy (PGP). (d) Email communication. (e) Yeast protein interactions. (f) PDZ domain interactions.}
\label{cor_bc_real}
\end{figure}
\begin{figure}[!]
\includegraphics[width=.5\columnwidth]{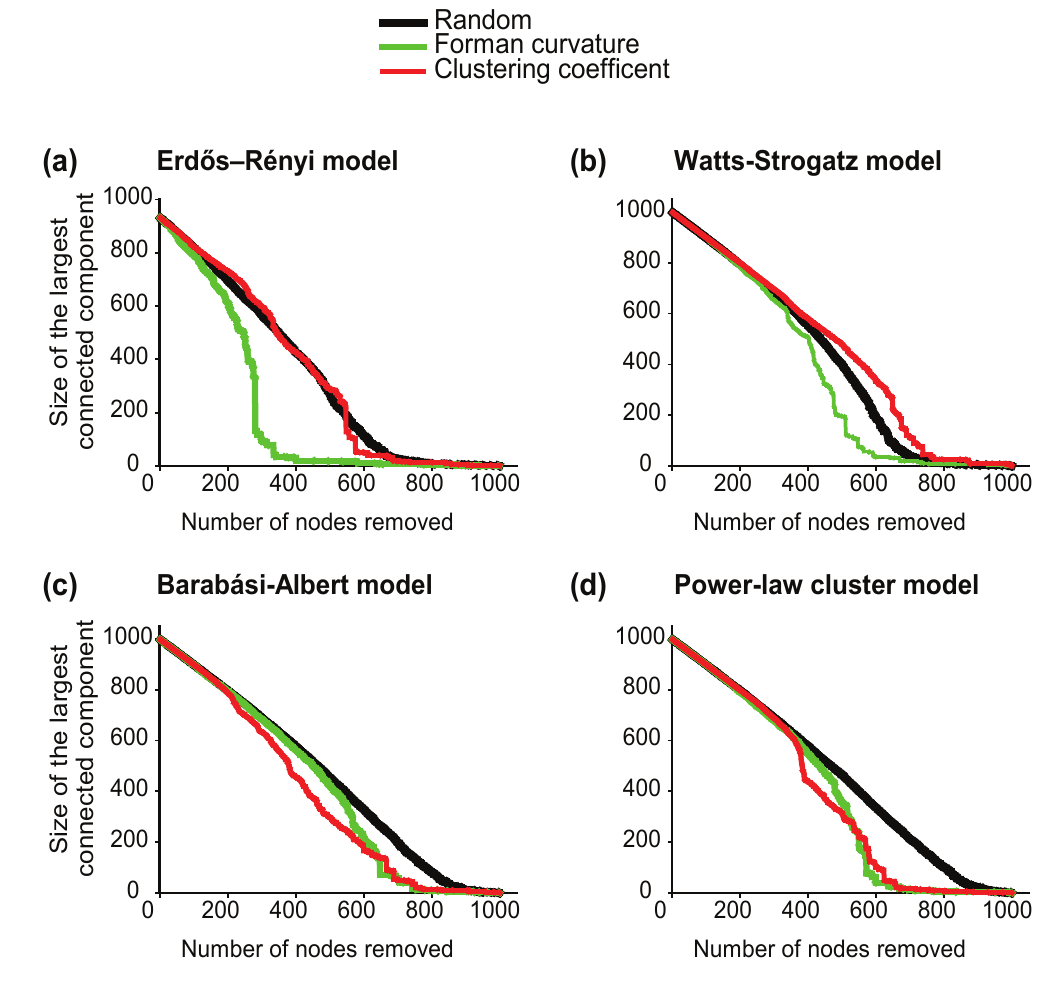}
\caption{Size of the largest connected component as a function of the number of nodes removed in model networks. (a) Erd\"{o}s-R\'{e}nyi (ER) model. (b) Watts-Strogratz (WS) model. (c) Barab\'{a}si-Albert (BA) model. (d) Power-law cluster (PLC) model. In this figure, the order in which the nodes are removed is based on the following criteria: Random order, Increasing order of Forman curvature, and Decreasing order of clustering coefficient. The parameters used to construct graphs from these generative models are same as those mentioned in Figure \ref{dist_model}.}
\label{rob_model}
\end{figure}

\begin{figure}[!]
\includegraphics[width=.5\columnwidth]{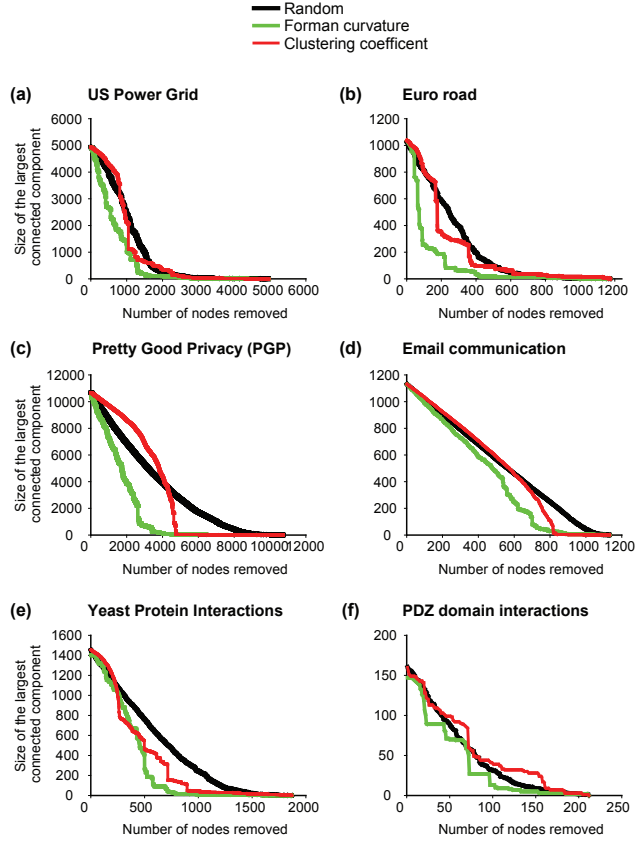}
\caption{Size of the largest connected component as a function of the number of nodes removed in real networks. (a) US Power Grid. (b) Euro road. (c) Pretty Good Privacy (PGP). (d) Email communication. (e) Yeast protein interactions. (f) PDZ domain interactions. In this figure, the order in which the nodes are removed is based on the following criteria: Random order, Increasing order of Forman curvature, and Decreasing order of clustering coefficient.}
\label{rob_real}
\end{figure}

\begin{figure}[!]
\includegraphics[width=.5\columnwidth]{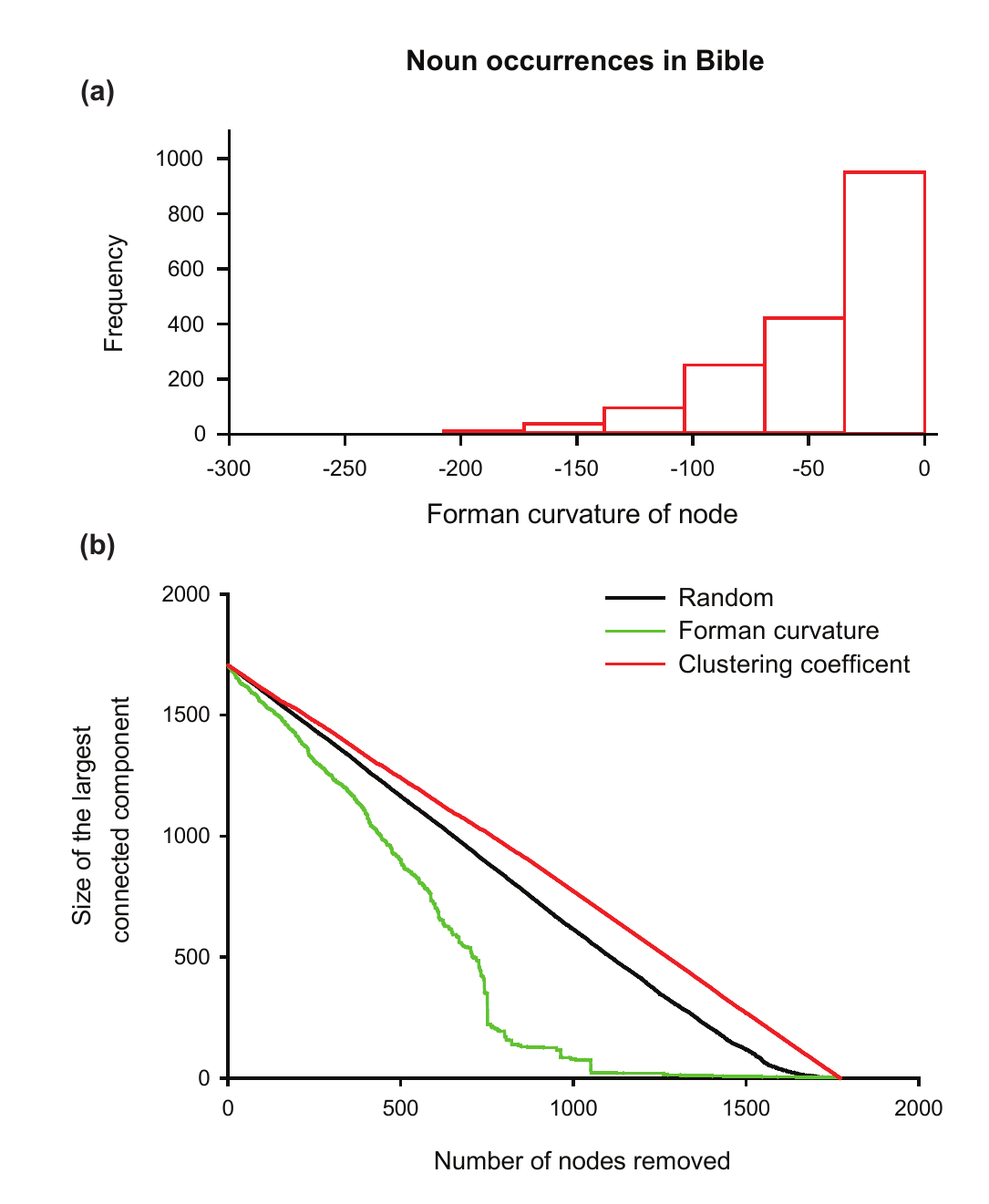}
\caption{Analysis of Forman curvature in a positively-weighted undirected network. (a) Distribution of Forman curvature of nodes in the network of Noun occurrences in Bible. (b) Size of the largest connected component as a function of the number of nodes removed in the network of Noun occurrences in Bible. The order of removing nodes is based on the following criteria: Random order, Increasing order of Forman curvature, and Decreasing order of clustering coefficient.}
\label{weighted}
\end{figure}

\section{Results}
\label{results}

\subsection{Distribution of Forman curvature in model and real networks}
\label{distribution}

Figure \ref{dist_model} shows the distribution of Forman curvature of nodes in model networks. It is seen that most nodes in model networks have negative curvature. Moreover, there is a clear difference in the nature of distribution of node curvature in different types of model networks. Specifically, in random networks generated by ER model and small-world networks generated by WS model, the distribution of node curvature is narrow with most nodes having curvature close to -4. But in scale-free networks generated by BA model and PLC model, the distribution of node curvature is broader with several nodes having curvature less than -40. In Supplementary Figure S1, we show the distribution of Forman curvature of edges in model networks. Similar to nodes, most edges in model networks have negative curvature. Also, there is a clear difference in the nature of distribution of edge curvature in different types of model networks. Random networks and small-world networks have a narrow distribution of edge curvature while scale-free networks have a broader distribution of edge curvature.

Figure \ref{dist_real} shows the distribution of Forman curvature of nodes in real networks. It is seen that most nodes in real networks have negative curvature. Moreover, the analyzed real networks have a broad distribution of node curvature. By comparing distributions of node curvature in model and real networks (cf. Figures \ref{dist_model} and \ref{dist_real}), it is clear that the nature of distribution in real networks is closer to that for scale-free networks. These findings are consistent with well-known result that most real-world networks have a scale-free architecture with power-law degree distribution \cite{Albert2002}. In Supplementary Figure S2, we show the distribution of Forman curvature of edges in analyzed real networks. Similar to model networks, the distribution of edge curvature in analyzed real networks is similar to the distribution of node curvature.

Figure \ref{PDZ_network} shows the network of PDZ-domain mediated protein-protein interactions \cite{Beuming2005}. It is seen that most nodes and edges in the PDZ network have negative curvature. In the PDZ network, high-degree nodes or hubs and central nodes or bottlenecks seem to have highly negative curvature.
PDZ network is disassortative with negative degree assortativity, and it can be seen that the high-degree nodes in this network are less likely to be directly connected to each other.

Figure \ref{Euroroad_network} shows the Euro road network \cite{Subelj2011} based on the international E-road network connecting cities located mostly in Europe. Similar to PDZ network, most nodes in the Euro road network have negative curvature. However, the Euro road network has positive degree assortativity unlike PDZ network, and it can be seen that the high-degree nodes in this network are more likely to be directly connected to each other.

In the classical Riemannian setting, Ricci curvature controls the growth of volumes (see for example \cite{Berger2000,Jost2011}). More precisely, spaces of negative Ricci curvature have exponential type growth, whereas those of positive Ricci curvature have a finite diameter. The same also holds for the Forman's discretization of the Ricci curvature \cite{Forman2003}. Since majority of the nodes and edges in considered networks essentially display a negative curvature, it follows that these networks display the potential of infinite growth. This important realization for real networks can be exploited for potential applications.


\subsection{Forman curvature and common network measures in model and real networks}
\label{correlation}

Degree of a node is given by the number of edges incident to the node in the network. Figure \ref{cor_degree_model} shows the correlation between degree and Forman curvature of nodes in model networks. We find a high negative correlation between node curvature and node degree in random networks generated by ER model and small-world networks generated by WS model. However, a weaker negative correlation is obtained between node curvature and node degree in scale-free networks generated by BA model and PLC model. Note that Forman curvature of a node is computed based on Forman curvature of edges incident on the node, and the Forman curvature of an edge in the network is dependent on its neighboring edges.

Figure \ref{cor_degree_real} shows the correlation between degree and Forman curvature of nodes in real networks. Interestingly, high negative correlation between node curvature and node degree is found in transportation and communication networks such as US Power Grid, Euro road and Email communication, while a weaker negative correlation between node curvature and node degree is found in biological networks such as Yeast protein interactions and PDZ domain interactions. Supplementary Table S5 summarizes the results of these analyses between Forman curvature and degree of nodes in model and real networks.

In Supplementary Figure S3, we plot the value of correlation between Forman curvature and degree of nodes in real networks as a function of degree assortativity of considered networks. This figure suggests that disassortative networks such as Yeast protein interactions and PDZ domain interactions have weaker correlation between node curvature and node degree. Moreover, communication and transportation networks with positive degree assortativity have high correlation between node curvature and node degree. In future, it will be worthwhile to expand the investigation of the observed relationship between degree assortativity and value of correlation between Forman curvature and degree of nodes to other networks.

Clustering coefficient \cite{Holland1971,Watts1998} of a node is given by the number of edges that are realized between the neighbors of the node divided by the number of edges that could possibly exist between the neighbors of the node in the network. Clustering coefficient quantifies the tendency of nodes to cluster together in a network. Figure \ref{cor_cc_model} shows the correlation between clustering coefficient and Forman curvature of nodes in model networks. Interestingly, no correlation is obtained between Forman curvature and clustering coefficient of nodes in random networks generated by ER model and small-world networks generated by WS model. However, a weak negative correlation is obtained between Forman curvature and clustering coefficient of nodes in scale-free networks generated by BA model and PLC model.

Since clustering coefficient is another measure to quantify curvature in complex networks \cite{Bridson1999,Eckmann2002}, we investigated the correlation between degree and clustering coefficient of nodes in model networks (cf. Supplementary Table S6). We find very weak or no correlation between degree and clustering coefficient of nodes in random networks generated by ER model and small-world networks generated by WS model. Moreover, we find a weak negative correlation between degree and clustering coefficient of nodes in scale-free networks generated by BA model and PLC model. However, the absolute value of the negative correlation between degree and clustering coefficient of nodes is lower than the absolute value of the negative correlation between degree and Forman curvature of nodes in scale-free networks (cf. Supplementary Tables S5 and S6).

Figure \ref{cor_cc_real} shows the correlation between clustering coefficient and Forman curvature of nodes in real networks. It can be seen that there is either extremely weak or no correlation between clustering coefficient and Forman curvature of nodes in considered real networks. Also, there is very weak or no correlation between degree and clustering coefficient of nodes in the analyzed real networks (cf. Supplementary Table S6). Notably, the absolute value of correlation between degree and Forman curvature of nodes is much higher in real networks compared to the observed correlation between degree and clustering coefficient of nodes in the same networks (cf. Supplementary Tables S5 and S6). Presumably, Forman curvature unlike clustering coefficient, has better correlation with degree because this type of curvature by its definition captures the geometric properties of the classical notions for smooth spaces discretized by networks.

Betweenness centrality \cite{Freeman1977,Newman2010} of a node gives the fraction of shortest paths between all pairs of nodes in the network that pass through that node. This measure is an indicator of the centrality of a node in the network, and a node with high betweenness centrality is a bottleneck for flows in the network. Figure \ref{cor_bc_model} shows the correlation between betweenness centrality and Forman curvature of nodes in model networks. Interestingly, high negative correlation is obtained between Forman curvature and betweenness centrality of nodes in random networks generated by ER model and small-world networks generated by WS model. Also, significant negative correlation is obtained between Forman curvature and betweenness centrality of nodes in scale-free networks generated by BA model and PLC model. Figure \ref{cor_bc_real} shows the correlation between betweenness centrality and Forman curvature of nodes in real networks. It can be seen that there is a weak to high negative correlation between betweenness centrality and Forman curvature of nodes in considered real networks.

We have also studied the correlation between Forman curvature and two other centrality measures, eigenvector centrality \cite{Bonacich1987} and closeness centrality \cite{Bavelas1950,Sabidussi1966}, in model and real networks. Similar to betweenness centrality, we find a high negative correlation between Forman curvature and closeness (or eigenvector) centrality of nodes in considered model networks (cf. Supplementary Table S5). Also, we find a clear negative correlation between Forman curvature and closeness (or eigenvector) centrality of nodes in analyzed real networks (cf. Supplementary Table S5).

Since clustering coefficient is another measure to quantify curvature in complex networks, we have also investigated the correlation between betweenness, closeness or eigenvector centrality and clustering coefficient of nodes in model and real networks (cf. Supplementary Table S6). We find in majority of considered model and real networks, there is no or weak correlation between betweenness, closeness or eigenvector centrality and clustering coefficient of nodes (cf. Supplementary Table S6). Moreover, there is a clear lack of consistent trend in the correlation between betweenness, closeness or eigenvector centrality and clustering coefficient of nodes in considered model and real networks (cf. Supplementary Table S6). Our results underscore that Forman curvature has stronger association with centrality measures compared to clustering coefficient.


\subsection{Forman curvature and topological robustness of networks}
\label{robustness}

We next investigated the effect of removing nodes based on increasing order of their Forman curvature on the large-scale connectivity of networks. Figure \ref{rob_model} shows the size of the largest connected component in model networks as a function of the number of nodes removed. Here, the order of removing nodes is based on the following criteria: (a) Random order, (b) Increasing order of Forman curvature (i.e, starting from the node with most negative curvature), and (c) Decreasing order of clustering coefficient (i.e, starting from the node with highest clustering coefficient). By comparing the effect of random removal against Forman curvature based removal in Figure \ref{rob_model}, it is clear that targeted removal of nodes with highly negative Forman curvature leads to faster disintegration compared to random removal of nodes in model networks. Moreover, it is found that removal of nodes based on increasing order of Forman curvature leads to faster disintegration compared to removal of nodes based on decreasing order of clustering coefficient in model networks.

Figure \ref{rob_real} shows the size of the largest connected component in real networks as a function of the number of nodes removed. Similar to model networks, we find that targeted removal of nodes with highly negative Forman curvature leads to faster disintegration compared to random removal of nodes in real networks. Also, we find that removal of nodes based on increasing order of Forman curvature leads to faster disintegration compared to removal of nodes based on decreasing order of clustering coefficient in real networks. Our results suggest that nodes with highly negative Forman curvature seem more important than nodes with high clustering coefficient to maintain the large-scale connectivity of model and real networks.

Earlier work \cite{Barabasi1999,Jeong2000,Jeong2001,Albert2002,Joy2005,Yu2007,Newman2010} has shown that diverse networks are vulnerable to targeted removal of nodes with high degree or high betweenness centrality. Thus, we decided to compare the effect of removing nodes based on increasing order of their Forman curvature against removing nodes based on decreasing order of their degree or betweenness centrality on the large-scale connectivity of networks (cf. Supplementary Figures S4 and S5). We find that the removal of nodes based on decreasing order of degree or betweenness centrality leads to faster disintegration compared to removal of nodes based on increasing order of Forman curvature in considered model and real networks (cf. Supplementary Figures S4 and S5). Thus, nodes with highly negative Forman curvature are more important than nodes with high clustering coefficient for maintenance of the overall connectivity of model and real networks, but such nodes with highly negative Forman curvature are less important compared to nodes with high degree or high betweenness centrality.


\subsection{Forman curvature and weighted networks}
\label{weight}

The definition of Forman curvature (Equation \ref{FormanRicciEdge}) elegantly incorporates the weights of nodes and edges in networks, and this feature renders Forman curvature an  attractive measure to study the organization of complex networks. Notice that the clustering coefficient unlike Forman curvature is unable to account for weights of nodes and edges in networks. So far, we have reported results based on our investigation of undirected and unweighted networks. Here, we report the Forman curvature in the network of Noun occurrences in Bible \cite{Basu2016} which is a undirected but positively-weighted network.

Figure \ref{weighted}(a) shows the distribution of Forman curvature of nodes in the network of Noun occurrences in Bible. It is seen that most nodes have a negative curvature but the distribution is broad with several nodes having highly negative curvature ($\le$ -100). Figure \ref{weighted}(b) shows the size of the largest connected component in the network of Noun occurrences in Bible as a function of the number of nodes removed. In Figure \ref{weighted}(b), the order of removing nodes is based on the following criteria: (a) Random order, (b) Increasing order of Forman curvature, and (c) Decreasing order of clustering coefficient. We find that targeted removal of nodes with highly negative Forman curvature leads to faster disintegration compared to random removal of nodes in this weighted undirected network. Also, the removal of nodes based on increasing order of Forman curvature leads to faster disintegration compared to removal of nodes based on decreasing order of clustering coefficient in this weighted undirected network.


\section{Summary and Conclusions}
\label{conclusion}

We have introduced here the Forman curvature, which represents a new type of discretization of Ricci curvature, to the realm of undirected networks. The mathematical formula of the Forman curvature (Equation \ref{FormanRicciEdge}) elegantly incorporates the weights of nodes and edges in networks. Hence, Forman curvature can be utilized to study the structure of both unweighted and weighted complex networks.

By investigating the distribution of Forman curvature of nodes and edges in model and real networks, we find that most nodes and edges in considered networks have a negative curvature. The distribution of node curvature and edge curvature is narrow in both random and small-world networks, while the distribution is broad in scale-free networks. In real networks, the distribution of node curvature and edge curvature is also broad like scale-free networks. Our results highlight that the distribution of Ricci curvature enables distinction between the different types of model networks (cf. Figure \ref{dist_model}), but also between the different types of real networks (cf. Figure \ref{dist_real}). This suggests that Forman curvature can be used to classify different types of networks.

We next investigated the correlation between Forman curvature and common network measures including degree, clustering coefficient and betweenness centrality, in model and real networks. High negative correlation is observed between Forman curvature and degree in both random and small-world networks, while a weaker negative correlation is seen in scale-free networks. Among real networks, a high negative correlation between Forman curvature and degree is found in transportation and communication networks such as US Power Grid, Euro road and Email communication, while a weaker negative correlation is found in biological networks such as Yeast protein interactions and PDZ domain interactions. Our extensive analysis of real networks also uncovers a possible relationship between degree assortativity and value of correlation between Forman curvature and degree. We also find a significant negative correlation between Forman curvature and different centrality measures including betweenness, eigenvector and closeness centrality in random, small-world and scale-free networks. In real networks, we find a weak to high negative correlation between Forman curvature and centrality measures.

Clustering coefficient has been used as a reference measure to quantify curvature in complex networks \cite{Bridson1999,Eckmann2002}. We find no correlation between Forman curvature and clustering coefficient in random and small-world networks, while a weak negative correlation was found in scale-free networks. Similarly, there is an extremely weak or no correlation between Forman curvature and clustering coefficient in analyzed real networks. Our results underline that Forman curvature has stronger association with degree and centrality measures compared to clustering coefficient.

By investigating the effect of removing nodes based on their Forman curvature on the large-scale connectivity of networks, we show that both model and real networks are vulnerable to targeted removal of nodes with highly negative Forman curvature. Moreover, in both model and real networks, node removal based on increasing order of Forman curvature leads to faster disintegration compared to node removal based on decreasing order of clustering coefficient. Our results underscore the importance of nodes with highly negative Forman curvature compared to those with high clustering coefficient for the maintenance of the large-scale connectivity in networks.

Forman's curvature for networks introduced here like the more established Ollivier's curvature \cite{Ollivier2009,Lin2010,Bauer2012,Jost2014,Ni2015,Sandhu2015a} for networks is a discretization of the the classical Ricci curvature. Previously, Ni \textit{et al} \cite{Ni2015} have studied the Ollivier's curvature in both model and real networks. Interestingly, we find that our results for Forman's curvature in networks are qualitatively similar to those obtained by Ni \textit{et al} \cite{Ni2015} for the Ollivier's curvature in networks. Like Ni \textit{et al} \cite{Ni2015} for Ollivier's curvature, in both model and real networks, we find that most nodes and edges have a negative Forman curvature, and are vulnerable to targeted deletion of nodes with negative Forman curvature. However, the analysis of Ni \textit{et al} \cite{Ni2015} for Ollivier's curvature was limited to communication networks among real networks. In contrast, we have performed an extensive analysis of the Forman curvature that spans across biological, social and communication networks. Fundamentally, still the qualitative agreement seen between our results for Forman's curvature and Ni \textit{et al} for Ollivier's curvature, reinforces the strength of both types of curvature for complex networks, and shows that the two curvatures, both in theoretical and practical sense, represent good discretizations of Ricci curvature.

Strikingly, the formula to compute Forman's curvature in networks is extremely simple and can easily scale to very large networks. In contrast, the calculation of Ollivier's curvature in networks is computationally intensive as it requires solving a difficult linear programming problem associated with the optimal mass transport. Thus, the computation of Ollivier's curvature may be computationally infeasible in extremely large networks. Also, a detailed comparative analysis of Forman's curvature and Ollivier's curvature in diverse networks is a natural follow-up of this work, but such follow-up analysis will require substantial more effort compared to this work.

Several other natural extensions of the present work can be foreseen in the near future. In the present work, we have ported the concept of Forman curvature from differential geometry to the sphere of undirected networks. However, several important real-world networks including the world wide web (WWW) \cite{Broder2000}, metabolic network \cite{Jeong2000}, gene regulatory network \cite{Milo2002} and neural network \cite{Sporns2004} are inherently directed in nature. An extension of the Forman curvature to directed edges is necessary for proper investigation of directed networks.

Notice that the mathematical formula of the Forman curvature (Equation \ref{FormanRicciEdge}) contains terms with square root of node weights and edge weights in the denominator, and thus, the formula of the Forman curvature can only be utilized to compute curvature in networks with positive weights of nodes and edges. We remark that the formula of the Forman curvature for edges in networks was derived from a more general formula by Forman which is applicable to a broad class of geometric objects, and thus, the weights in the present formula correspond to the Riemannian notions of length, area, volume, etc., considered by Forman which must be positive. However, nodes and edges in real-world networks such as gene regulatory network \cite{Milo2002} and neural network \cite{Sporns2004} can have negative weights. For example, negative edge weights in gene regulatory network represent repressive interactions between transcription factors and regulated genes. Thus, an extension of the Forman curvature to account for negative weights of edges and nodes will enable proper investigation of signed networks.

We conclude with a suggestion for another possible extension that stems from a theoretical, albeit straightforward, reflection of the mathematical formula of the Forman curvature. As mentioned in section \ref{FormanCurvature}, the definition of the Forman curvature is coupled with two fitting Laplacians. In future, it will be interesting to utilize these Laplacians associated with the Forman curvature for standard applications of spectral analysis and denoising in large-scale networks.


\begin{acknowledgments}
ES and AS thank the Max Planck Institute for Mathematics in the Sciences, Leipzig, for their warm hospitality. AS acknowledges support from Max Planck India Mobility Grant, IMSc PRISM project, Ramanujan fellowship (SB/S2/RJN-006/2014) and Department of Science and Technology (DST) start-up project (YSS/2015/000060). ES would like to thank Jie Gao, David Gu, and especially, Eli Appleboim, for interesting discussions.
\end{acknowledgments}

\appendix
\section{Interpretation of Forman curvature}

\begin{figure*}
\includegraphics[width=10cm]{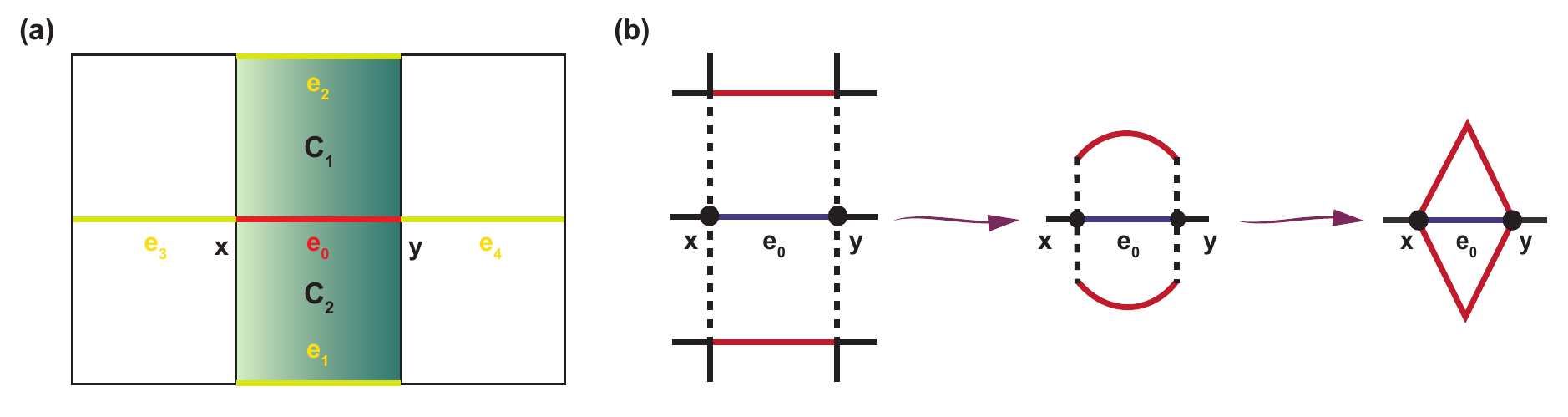}
\caption{The role of parallel edges in Forman's curvature. (a) Simple case of a square grid. The edges parallel to $e_0$ have in common with it either a \textit{child} ($e_1$ and $e_2$), or a \textit{parent} ($e_3$ and $e_4$). (b) The extrinsic length of an arc $\widehat{xy}$ compared to the intrinsic one given by the length of the segment $\overline{xy}$, and the role of parallel edges, as illustrated by the degeneration of a square grid to a triangular one.}
\label{curvature}
\end{figure*}

As mentioned in section \ref{FormanCurvature}, Forman's discretization of Ricci curvature is derived from the so called Bochner-Weitzenb\"{o}ck formula, and, as such, its geometric content is, perhaps, less than transparent. However, we can still gain some insight into this curvature measure by following Forman's own approach \cite{Forman2003}. Forman's curvature was devised as a discrete version of the classical Ricci curvature by keeping in mind surfaces and higher dimensional (smooth) manifolds. Still one can gain intuition on Forman's curvature by considering the simpler yet relevant case of two-dimensional surfaces.

A first important insight to gain arises from the fact that Ricci curvature is \textit{intrinsic}. An \textit{intrinsic} measure is independent of the \textit{embedding} in the space, while an \textit{extrinsic} measure is dependent upon the chosen (or given) embedding. In the smooth setting, isometric embeddings are formally defined as follows. Let $M^m, N^n$ be smooth Riemannian manifolds and let $f:M^m  \rightarrow N^n$ be a smooth mapping. Then $f$ is called:
\begin{itemize}
\item an {\em immersion} if $df$ has maximal rank (i.e., equal to $m$) at every point $p \in M^m$.
\item an {\em embedding} if it is an immersion and a homeomorphism onto its image.
\item an {\em isometric embedding} if it is an embedding and if it preserves the length of (smooth) curves.
\end{itemize}
Intuitively, a property (e.g. length, curvature, etc.) is intrinsic if it does not depend upon the specific realization of the given geometrical object (e.g. of a surface in Euclidean space). In our context, it means that Forman's curvature depends solely on the combinatorial data (i.e., adjacency matrix) and prescribed weights of the network. But Forman's curvature is not dependent, for instance, on the way the edges are drawn, say in the plane, and nor on the relative magnitudes or distribution of the weights. Since in Forman's original concept, edge weights represent abstractizations of the notion of length, the connection with the classical notion of intrinsic embedding becomes quite transparent (see also the discussion below).

Let us now see how this definition translates into a more discrete setting by considering a square grid shown in Figure \ref{curvature}(a).  For the square grid in Figure \ref{curvature}(a), the following points need to be emphasized. First, in the classical (smooth manifold) setting, Forman-Ricci curvature is measured along a vector whose discrete analogue is an edge. Second, the intrinsic \textit{distances} on the square grid shown in Figure \ref{curvature}(a) need to be identified. Following Forman \cite{Forman2003}, the only edges contributing to the intrinsic \textit{distances} are the ones parallel to the edge $e_0$ under consideration in Figure \ref{curvature}(a). That is, those edges that either have a common \textit{parent} (higher dimensional face) or a common \textit{child} (lower dimensional face) with edge $e_0$ contribute to the intrinsic \textit{distances}, but not those edges which have both a common parent and common child. One can convince themselves about the truth of this assertion by passing to a degenerate case, namely, by collapsing the \textit{incoming} (adjacent) edges as shown in Figure \ref{curvature}(b). It can be shown that the parallel edges are precisely the ones that make a contribution to the meaningful Laplacian which is intrinsically associated with the Forman-Ricci curvature along the edge $e_0$ (for details see \cite{Forman2003}). Thus, based on these observations, one can regard the Forman-Ricci curvature of an edge as a measure of the weighted flow along the given edge.

In the limiting case of graphs or networks considered here, there are no higher dimensional faces than the one-dimensional faces which correspond to the edges of the network, and this realization leads to the mathematical formula (Equation \ref{FormanRicciEdge}) for the Forman curvature of the edge. We remark that the interpretation of the Forman-Ricci curvature still holds for the case of networks. Namely, the Forman curvature is a measure of the flow across an edge in the network, as contributed by the edges adjacent to it.



%

\end{document}